\documentclass[twocolumn]{revtex4}

\usepackage{lineno,hyperref}

\usepackage{graphicx}
\usepackage{amssymb}
\usepackage{amsmath}
\usepackage{latexsym}
\usepackage{ulem}
\usepackage{color}
\usepackage{dcolumn}
\usepackage{subfigure}
\usepackage[T1]{fontenc}
\usepackage[utf8]{inputenc}
\usepackage[english,frenchb]{babel}
\usepackage{physics}
\usepackage{amsmath}
\let\oldAA\AA
\renewcommand{\AA}{\text{\normalfont\oldAA}}

\begin{document}

\title{Dramatic effects of vacancies on phonon lifetime and thermal conductivity in graphene}

\author{G. Bouzerar$^1$, S. Thébaud$^1$, S. Pecorario$^1$ and Ch. Adessi$^1$
\\
$^1$ CNRS et Université Lyon 1, 6, rue Ada Byron, 69622 Villeurbanne Cedex, France\\
}
\date{\today}
\selectlanguage{english}

\begin{abstract}
Understanding thermal transport in 2D materials and especially in graphene is a key challenge for the design of heat management and energy conversion devices. The high sensitivity of measured transport properties to structural defects, ripples and vacancies is of crucial importance in these materials. Using a first principle based approach combined with an exact treatment of the disorder, we address the impact of vacancies on phonon lifetimes and thermal transport in graphene. We find that perturbation theory fails completely and overestimates phonon lifetimes by almost two orders of magnitude. Whilst, in defected graphene, LA and TA modes remain well defined, the ZA modes become marginal. In the long wavelength limit, the ZA dispersion changes from quadratic to linear and the scattering rate is found proportional to the phonon energy, in contrast to the quadratic scaling often assumed. The impact on thermal transport, calculated beyond the relaxation time approximation and including first principle phonon-phonon scattering rates as reported recently for pristine graphene, reveals spectacular effects even for extremely low vacancy concentrations.
\end{abstract}

\maketitle

\section{Introduction}

Over the past decade, and because of its unique mechanical, electronic, optical and thermal properties, the 2D material prototype, graphene, has been at the heart of a plethora of publications.\cite{review1,review2,review3} Among all these remarkable properties, the particular topic of thermal transport has attracted much attention over the past years. Graphene exhibits an unusually high thermal conductivity,\cite{balandin1,gosh1,cai,chen} that could, in the near future, make this two dimensional material one of the best candidates for efficient thermal dissipation in microelectronics. In addition, nanostructuring \cite{prasher} and disorder \cite{cuniberti} in graphene based compounds could be promising pathways for high-efficiency thermoelectric devices to transform waste heat into electrical energy. The desired effects are a drastic suppression of the thermal conductivity and significant improvement of the Seebeck coefficient, both required to reach a large thermoelectric efficiency.

Many efforts are still devoted to understanding the lattice thermal conductivity in graphene, but no clear consensus has been reached so far, the topic remains controversial. Experimentally, the measured room temperature thermal conductivity ($\kappa$) spans over a large range of values, typically between 400 W.m.$^{-1}$K$^{-1}$ and 600 W.m.$^{-1}$K$^{-1}$ for supported samples\cite{cai,seol} and 1500 W.m.$^{-1}$K$^{-1}$ to 5400 W.m.$^{-1}$K$^{-1}$ for suspended ones.\cite{balandin1,gosh2,cai} The important fluctuations in the measured values can be attributed to (i) large measurement uncertainties, (ii) variations in the processing conditions and (iii) graphene quality. This could indicate a high sensitivity to intrinsic defects such as vacancies, lattice reconstruction, edge roughness or even ripples. From the theoretical side, \cite{weerasinghe,xie,fugallo,pauletto,lindsay,evans} the scenario is even more open. The estimates of $\kappa$ at room temperature in graphene vary by more than one order of magnitude, typically it ranges from 500 W.m.$^{-1}$K$^{-1}$ to about 9000 W.m.$^{-1}$K$^{-1}$.

In this paper, using state of the art first principle based approaches, we address the issue of vacancies' effects on lattice thermal transport in single mono-layer graphene. For that purpose, we follow a two steps procedure. First, we calculate the vacancy induced multiple scattering contribution to the phonon lifetime by an exact real space treatment of the disorder. Notice that, in most of the existing theoretical studies, disorder is treated perturbatively. As will be seen, the second order perturbation theory appears to severely overestimate the phonon lifetimes. Note however, that the perturbation theory often used in the case of isotopic disorder is reasonable, because of the weak effects of the substitution of $^{12}$C by $^{13}$C. In the second step, we calculate the thermal conductivity as a function of the vacancy concentration by including phonon-phonon scattering (Normal and Umklapp processes) and going beyond the relaxation time approximation (RTA). It should be emphasized that all calculations are parameter free, since the disordered phonon dynamical matrix is obtained from first principle calculations.

\begin{figure*}[t]\centerline
{\includegraphics[width=1.\linewidth,angle=0]{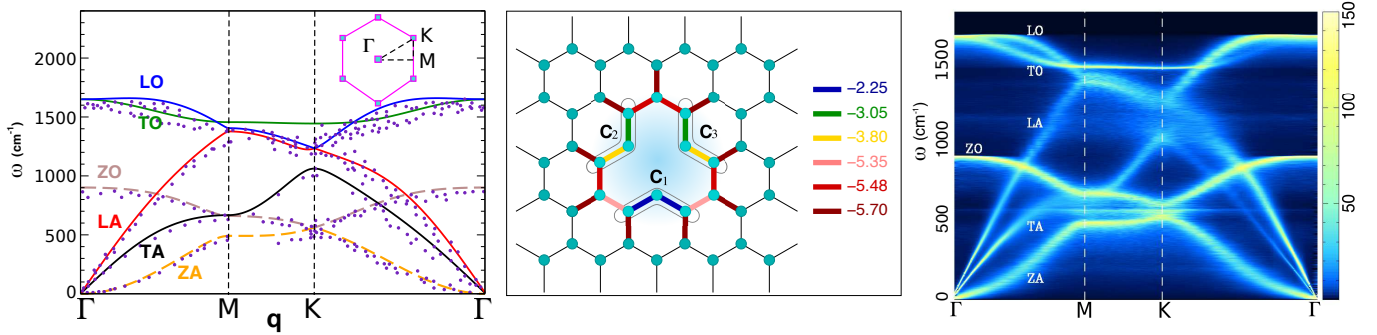}}
\caption{(left) Calculated phonon dispersions in pristine graphene along the $\Gamma$-M-K-$\Gamma$ path (continuous lines) for LA, TA, LO, TO, ZA and ZO branches. The dotted symbols are experimental data from Ref.\cite{wirtz,mohr,maultzsch}. (middle) {\it ab initio} calculated nearest-neighbour inter-atomic force constants for the ZA and ZO modes in the vicinity of a vacancy (in eV/${\AA}^{2}$). (right) 2D color plot of the calculated averaged phonon dynamical spectral function $A_{s}(\textbf{q},\omega)$ in the presence of 5$\%$ of vacancies. The chosen polarisation for the in-plane modes is $\textbf{e}_{s}=\frac{1}{\sqrt{2}}(\textbf{e}_x + \textbf{e}_y)$.}
\label{fig1}
\end{figure*}

\section{Exact disorder contribution to the phonon lifetime}

Fig.~\ref{fig1}(left) shows the calculated dispersion in the pristine graphene monolayer along the $\Gamma$-M-K-$\Gamma$ path in the Brillouin Zone (BZ). The inter-atomic force constants (IFC) $\phi_{\textbf{ij}}^{\alpha\beta}$ have been calculated from first principle simulations, $\textbf{i}$ and $\textbf{j}$ are the positions of the C atoms and $\alpha$,~$\beta$~=~ x, y and z. The {\it{ab initio}} calculations of the IFC are performed with the DFT package SIESTA,~\cite{siesta} more details are provided in the Appendix. Note that in the case of graphene, in-plane (xy) and out of plane (z) modes are decoupled. We observe an excellent agreement between our calculations and the experimental data, over the whole BZ. We should emphasize that, we have retained the first 6 nearest neighbour shells for the calculations, including further shells has negligible effects. 

In Fig.~\ref{fig1}(middle) are plotted the nearest-neighbour inter-atomic force constants $\phi_{n-n}^{zz}$ for the out of plane modes (ZA and ZO) in the vicinity of a vacancy. The rectangular supercell used for the calculations contains 199 C atoms. In a pure graphene mono-layer $\phi_{n-n}^{zz}$ is -6.1 eV/${\AA}^{2}$.
When a vacancy is introduced, $\phi_{n-n}^{zz}$ is strongly affected. More precisely, for the 2-coordinated C atoms the average value is only -3.03 eV/${\AA}^{2}$. The origin of the $\phi_{n-n}^{zz}$ variations between the 2-coordinated atoms is discussed in the Appendix.
As we move away from the vacancy we rapidly recover the value in the pristine compound. For larger distances between C atoms (beyond nearest-neighbours), the IFC are less affected by the presence of the vacancy.
In what follows, and for simplicity, the inclusion of a vacancy is treated as follows: (i) a removal of a C atom and the corresponding bonds and (ii) a reduction by a factor 2 of the nearest neighbour IFC of the 2-coordinated C atoms, for both in and out of plane modes.
We should mention, in contrast to what has been reported recently in a semi-empirical theoretical study,\cite{xie} that the $\phi_{n-n}^{\alpha\beta}$ are not enhanced by a factor 2 but strongly suppressed as revealed by our first principle calculations.

To obtain the vacancy induced multiple scattering contribution to the phonon scattering rate, we evaluate numerically on large systems (typically 10$^{6}$ C atoms) the phonon dynamical spectral function. Note that this function is directly accessible from inelastic neutron scattering experiments. For a fixed concentration of vacancies and given configuration of disorder (random positions of the vacancies), the dynamical spectral function reads (further details are available in the Appendix),
\begin{eqnarray}
A_{s}(\textbf{q},\omega)= -\frac{2}{\pi} \omega \Im (G_{s}(\textbf{q},\omega)),
\label{Aqw}
\end{eqnarray}
where
\begin{eqnarray}
G_{s}(\textbf{q},\omega)= \lim_{\eta \to 0}  \bra{\textbf{q},s}\Big[(\omega^{2} + i\eta) \hat{\bf 1} -\hat{\bf D}\Big]^{-1} \ket{\textbf{q},s},
\label{GF}
\end{eqnarray}
where $\hat{\bf D}$ is the dynamical matrix and $\ket{\textbf{q},s}$ is a Bloch state with momentum $\textbf{q}$ and polarisation vector $\textbf{e}_{s}$, and $\eta$ is a small imaginary part.
The matrix element $D^{\alpha\beta}_{\textbf{i}\textbf{j}}= \frac{\phi^{\alpha\beta}_{\textbf{i}\textbf{j}}}{\sqrt{m_{\textbf{i}}m_{\textbf{j}}}}x_{\textbf{i}}x_{\textbf{j}}$, where $x_{\textbf{i}}=0$ if the site $\textbf{i}$ is occupied by a vacancy, otherwise $x_{\textbf{i}}=1$. In addition, for 2-coordinated atoms $\phi^{\alpha\beta}_{\textbf{i}\textbf{j}}$ is half of its value in the pristine compound for nearest-neighbour IFC only. For a given $\textbf{q}$ of the BZ, the peaks in $A_{s}(\textbf{q},\omega)$ provide both the energy of the phonon modes and the lifetimes that correspond to the inverse of the full width at half maximum (FWHM) of the peaks. To extract reliably and accurately the phonon peak positions and more particularly the FWHM that are strongly $\eta$ dependent for low energy modes, the calculations are performed using the powerful iterative Chebyshev Polynomial Green's Function approach (CPGF). \cite{mucciolo,weisse} 
A similar methodology involving Chebyshev Polynomials to expand the time evolution operator has been developed earlier to address the dc conductivity in quasi crystals.\cite{mayou}
Note that performing a direct exact diagonalization of the disordered dynamical matrix, would require a large amount of both memory and CPU time. In particular the CPU time scales as N$^{3}$, where N is the total number of C atoms. In contrast, the calculation using CPGF scales linearly with the system size and the memory required is small, because there is no need to store large matrices. We should emphasize that, within CPGF the treatment of the disorder is exact. Thus, quantum interferences and localization phenomena are fully included.

Fig.~\ref{fig1}(right) represents the disorder averaged dynamical spectral function for both in plane and out of plane modes and a concentration of vacancies set to $x=5\%$. Note that, because of the large system sizes considered here, a few configurations of disorder are enough to get reliable statistical average. In addition, in this figure, a small finite $\eta$ has been kept for the sake of visibility. However, in what follows and in order to properly extract the exact disorder contribution to the phonon lifetime the limit $\eta \rightarrow 0$ will be properly carried out. In the vicinity of the $\Gamma$ point the phonon modes are well defined, but the ZA modes appear to have a much broader width than that of the in-plane acoustic modes. As we move away, along the $\Gamma$-M or $\Gamma$-K path, first the phonon width of the acoustic modes increases significantly (lifetime reduces) and then becomes almost constant. The case of the optical modes is slightly different. In particular, for TO branch, the width first increases rapidly then reaches a maximum and as we approach the zone boundary it decreases again. Along the M-K path, ZA, ZO and especially TO modes are very sharp in contrast to LA, TA and LO modes. In this region, it is difficult to separate LO and LA modes. It is worth mentioning two other interesting features: a well defined vacancy induced flat band located at about 500 cm$^{-1}$ and the disappearance (fuzzy region) of the phonon modes around 1200 cm$^{-1}$ in the vicinity of the K point.
\begin{figure}[t]\centerline
{\includegraphics[width=1.\linewidth ,angle=0]{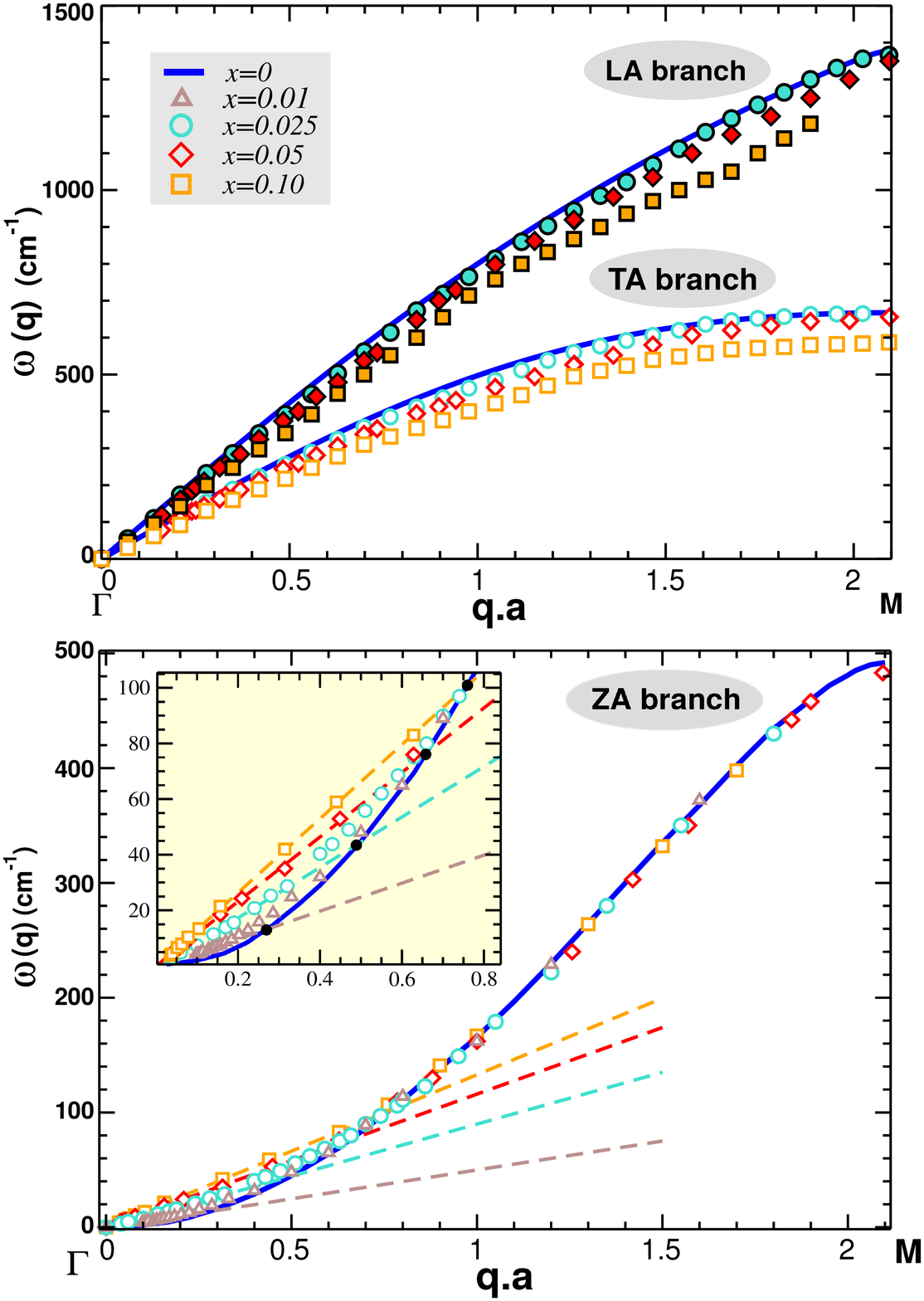}}
\caption{(top) LA and TA phonon dispersions along the $\Gamma$-M direction for various concentrations of vacancies ranging from $x=$ 0 to 0.1. (bottom) ZA phonon dispersion in the $\Gamma$M direction for the same concentrations. The dashed lines correspond to linear fits for the long wavelength phonons. The inset shows a zoom of this region.}
\label{fig2}
\end{figure}

Let us now discuss the effects of the vacancy concentration ($x$) on phonon dispersions. We choose here to focus our attention on the $\Gamma$-M direction. The results for the acoustic modes LA, TA and ZA are depicted in Fig.~\ref{fig2}. We first observe that both dispersions and velocities of LA and TA modes are weakly affected, the effects start to be visible for relatively large vacancy concentration of the order of 10$\%$. In contrast, the situation is very different for the ZA branch. For large values of the momentum $\vert\textbf{q}\vert$, the dispersion is insensitive to the vacancy concentration. However, in the vicinity of the $\Gamma$ point the dispersion goes from quadratic to linear. The long wavelength modes develop a finite velocity $v_{ZA}$ that depends on $x$. For instance, if we set $x=5\%$, we find that $v_{ZA}$ is as large as $0.15\, v_{LA}$. As clearly shown in the inset, the region of linear dispersion rapidly increases with the concentration of defects.

\begin{figure}[t]\centerline
{\includegraphics[width=1.\linewidth, angle=0]{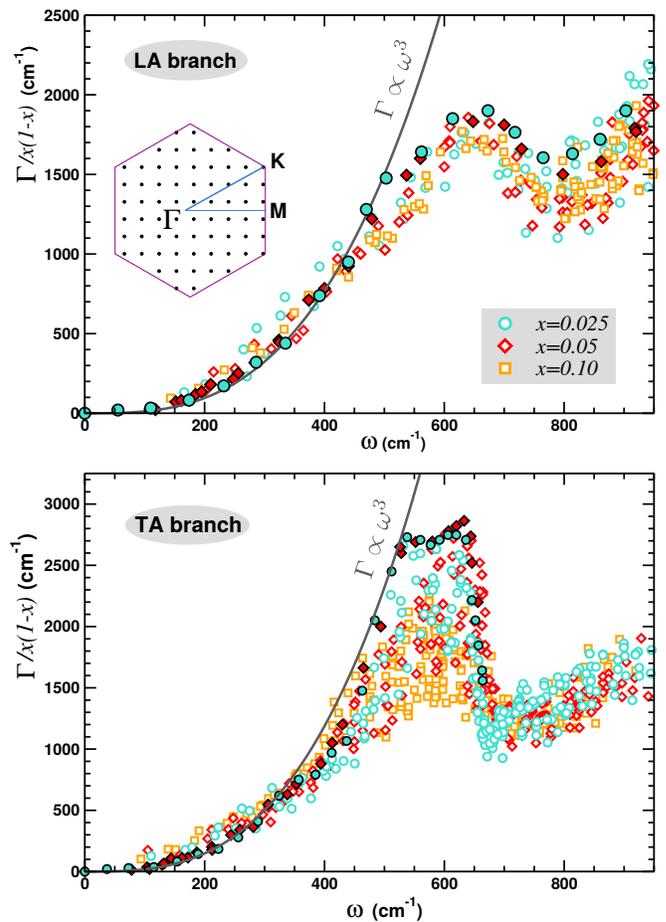}}
\caption{(top) LA phonon full width at half maximum (rescaled by $x(1-x)$) at $T=0$ K as a function of the phonon energy $\omega$ for $x=$ 0.025, 0.05 and 0.1. A grid of 200 to 300 $\rm \bf q$-points uniformly distributed over the whole Brillouin zone has been used. The filled symbols correspond to highly accurate calculations in the $\Gamma$-M direction (see text). The continuous line is a cubic fit of this set of data up to 500 cm$^{-1}$. (bottom) Same as in the top figure but for TA branch.
}
\label{fig3}
\end{figure}

We now propose to analyse the effects of vacancies on the phonon modes FWHM denoted $\Gamma (\omega)$. $A_{s}(\textbf{q},\omega)$ has been calculated in the whole BZ. Typically 200 to 300 $\textbf{q}$-points uniformly distributed over the BZ have been targeted. Note that the number of Chebyshev polynomials (CP) considered were typically of the order of 2.10$^{4}$ for $\textbf{q}$-points far from the BZ center. However, for the acoustic modes in the vicinity of the BZ center, because of the tiny values of the FWHM, it was necessary to include up to 3.10$^{6}$ CP to get converged results denoted "highly accurate calculations". $\Gamma(\omega)$ (rescaled by $x(1-x)$) as a function of the phonon mode energy is plotted in Fig.~\ref{fig3} for both LA and TA branches. The highly accurate calculations for the low energy modes in the $\Gamma$M direction are also shown. First, we observe for both branches a non monotonic behaviour of $\Gamma(\omega)$. We also find for $\omega \le 400$ cm$^{-1}$ that the data points obtained for various vacancy concentrations lie on the same curve. Beyond $400$ cm$^{-1}$, we observe, for a given energy, larger fluctuations around the average value. They are relatively small for the LA branch, of the order of 10 $\%$. In contrast, they are much larger for the TA modes. For instance, for $\omega = 600$ cm$^{-1}$ the fluctuations are of the order of 30 $\%$.
However, beyond $700$ cm$^{-1}$ the fluctuations are strongly suppressed. In addition, for $\omega \le 500$ cm$^{-1}$, the FWHM is found cubic in energy: $\Gamma_{\lambda} (\omega) = x(1-x) \frac{\omega^{3}}{\omega^{2}_{0,\lambda}}$ where $\omega_{0,\lambda} = $ 288 cm$^{-1}$ and 236 cm$^{-1}$ respectively for $\lambda =$ LA and TA.
The cubic power law found here is in agreement with perturbation theory (PT) that gives $\Gamma^{PT}_{\lambda}(\omega) = x \frac{\pi}{2}(\frac{\Delta M}{M})^{2} \omega^{2} \rho_{\lambda}(\omega)$, $\Delta M$ is the mass variation of the substituted atom and $\rho_{\lambda}(\omega)$ denotes the phonon density of states in the pristine compound.\cite{tamura}
In the particular case of vacancies $\frac{\Delta M}{M} = -\frac{M_{a}}{M} -2 \approx -3$, M$_{a}$ being the mass of the missing atom and M the average mass per atom, -2 accounts for the potential energy of the missing linkage. \cite{klemens1,klemens2} For LA and TA modes the density of states is  $\rho_{\lambda}(\omega) = \frac{\Omega}{2\pi v_{\lambda}^{2}} \omega$  where $\Omega$ is the primitive cell area. The comparison between PT and our exact results leads to $\Gamma^{PT}_{LA} \approx 0.018 ~\Gamma_{LA}$ and $\Gamma^{PT}_{TA} \approx 0.03 ~\Gamma_{TA}$.
Thus, perturbation theory drastically underestimates the vacancy contribution to the scattering rate. This is not surprising considering the non-perturbative nature of such defects compared to isotopic disorder. Note that in a recent study\cite{malekpour} devoted to thermal transport in irradiated graphene, a value of $(\frac{\Delta M}{M})^{2}$ of the order of 590 instead of 9 was found necessary to reproduce the experimental data. This is entirely consistent with our finding that LA scattering rate is about 55 times larger than that predicted by PT.

\begin{figure}[t]\centerline
{\includegraphics[width=1.\linewidth ,angle=0]{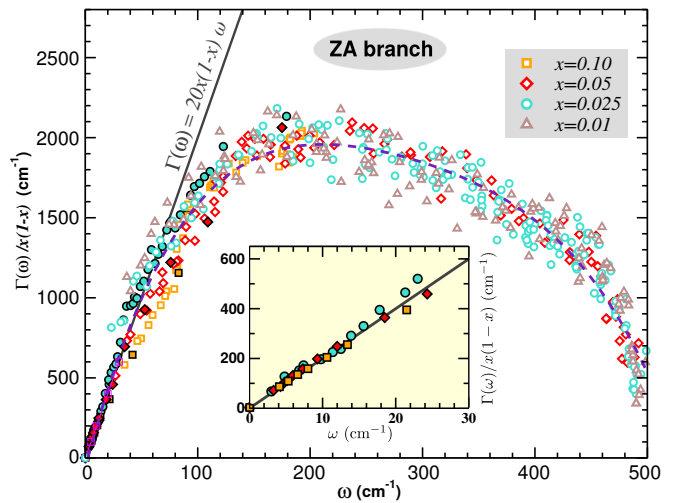}}
\caption{ZA phonon full width at half maximum (rescaled by $x(1-x)$) at $T=0$ K as a function of the phonon energy $\omega$ for $x =$ 0.01, 0.025, 0.05 and 0.1. A grid of 200 to 300 $\rm \bf q$-points uniformly distributed over the whole Brillouin zone has been used. The filled symbols correspond to highly accurate calculations in the $\Gamma$-M direction (see text). The continuous line is a linear fit for the low energy modes (see inset).}
\label{fig4}
\end{figure}

The case of the ZA branch is even more interesting. $\Gamma(\omega)$ for the ZA branch is plotted in Fig.~\ref{fig4}. First, after rescaling, the data points lie on a single curve as seen previously for in plane modes. The behaviour is non monotonic, $\Gamma (\omega)$ exhibits a maximum at about $200$ cm$^{-1}$ and a strong decrease as we approach the BZ boundary (M and K points). The fluctuations for a given $\omega$ are found relatively small, at most of the order of $10\%$ to $15\%$ around the average value.
Unexpectedly, as we introduce vacancies, we find a linear behaviour of $\Gamma_{ZA}$ as a function of the mode energy in the vicinity of the BZ center. The fluctuations are very small in this region as clearly seen in the inset. A linear fit for $\omega \le 100$ cm$^{-1}$ leads to $\Gamma_{ZA} (\omega) = 20 \, x(1-x)\,\omega$.
This linear scaling of the ZA width in the long wavelength regime means that these modes are marginal since $\lim_{\omega \to 0} \frac{\Gamma_{ZA}}{\omega} =  20 \, x(1-x)$. For well defined excitations, one expects this ratio to be zero.
Furthermore, $\Gamma_{ZA} (\omega) \ge \omega$ for $x \ge 0.05$, hence the phonon quasi-particles are not well defined in this range of vacancy concentrations. These findings are in strong contrast with perturbation theory. Indeed, it is expected that $\Gamma^{PT}_{ZA}(\omega) = x (\frac{\Delta M}{M})^{2} \frac{\Omega}{8 D} \omega^{2}$ where D is the stiffness of the ZA branch in the pristine compound, i.e. $\omega(\textbf{q}) =D \vert \textbf{q}\vert^2$ in the long wavelength limit. Therefore, the prediction of a quadratic power law from perturbation theory is inconsistent with the exact numerical results. For example, for $\omega=$ 50 cm$^{-1}$ and 100 cm$^{-1}$ we find $\frac{\Gamma^{PT}_{ZA}}{\Gamma_{ZA}}= 0.045$ and 0.09 respectively. Once again, perturbation theory severely overestimates the phonon lifetime and hence the mean free path of the ZA phonon modes in the presence of vacancies. Because the thermal conductivity is dominated by the ZA modes in both pristine graphene and in the presence of C isotopes,\cite{lindsay2} we naturally expect in the calculation of $\kappa$ a strong deviation from PT as vacancies are introduced.

\section{Effects of vacancies on the thermal conductivity}

Linearized Boltzmann Transport Equation (BTE) \cite{peierls,ziman} is a frequently used and efficient theoretical approach to address the thermal conductivity in 2D and 3D materials. In the great majority of studies, and because solving exactly the BTE is more cumbersome, \cite{omini1,omini2} the relaxation time approximation (RTA) is often assumed.
However, comparative studies have revealed that RTA calculated thermal conductivity is often much smaller than the full BTE solution
in 3D. \cite{broido1,esfarjani,ward,carrete} The discrepancy is even stronger in 2D compounds such as pristine or natural graphene monolayer. \cite{lindsay1,lindsay2,fugallo} The origin of the discrepancy is the fact that within RTA, Umklapp (U) and Normal (N) phonon-phonon scattering processes are treated on equal footings, as resistive. However, N processes are not resistive, the thermal conductivity is expected to diverge in the absence of U scattering. There is an alternative to full BTE that can be implemented more easily and corrects the shortcomings intrinsic to the RTA approach. Several decades ago, Callaway proposed a theory that allows U and N processes to be treated separately. \cite{callaway,allen,ma} It has been found that the Callaway theory leads to thermal conductivities that agree very well with the full BTE approach in 3D systems such as Si \cite{ward1} and even in graphene. \cite{cepellotti}
\begin{figure}[t]\centerline
{\includegraphics[width=1.\linewidth ,angle=0]{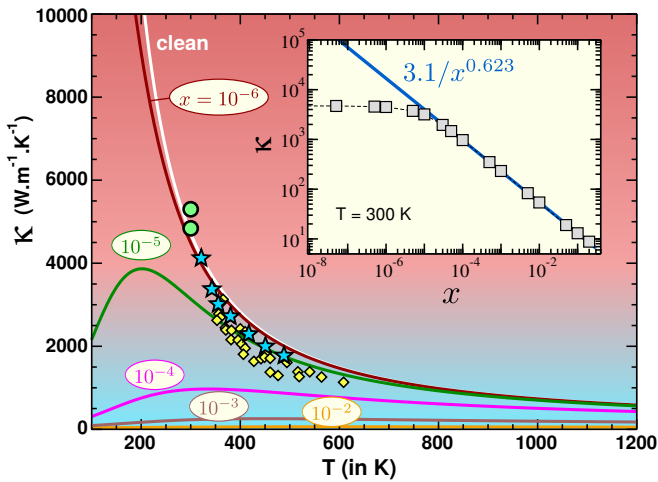}}
\caption{Thermal conductivity $\kappa$ in W.m.$^{-1}$K$^{-1}$ as a function of temperature for various concentrations of vacancies ranging from 0 to 0.01 (continuous lines).
The symbols (circles, stars and diamonds) are experimental measurements extracted from Refs.\cite{balandin1,nika,chen}. The inset represents the variation of the room temperature $\kappa$ as a function of $x$, the blue continuous line is a fit for $x\ge 10^{-5}$.
}
\label{fig5}
\end{figure}
Here, we propose to address the effects of the vacancies on thermal conductivity using the Callaway approach (see details in the Appendix. Within this approach the thermal conductivity in the $\alpha$-direction reads,
\begin{eqnarray}
\kappa^{\alpha}=\kappa^{\alpha}_{RTA} + \Delta\kappa^{\alpha},
\label{kappa}
\end{eqnarray}
the first term is the RTA contribution and the second one is the correction due to the appropriate separation between resistive and non-resistive processes. $\kappa^{\alpha}_{RTA}$ is given by,
\begin{eqnarray}
\kappa_{RTA}^{\alpha}=\frac{1}{k_{B}T^{2}}\frac{1}{N\Omega\delta} \sum_{\textbf{q},\lambda} (\hbar \omega_{\lambda})^{2}(v^{\alpha}_{\lambda})^{2} \times \nonumber \\
\tau^{tot}_{\lambda}f_{\lambda}^{0}(f_{\lambda}^{0}+1),
\label{kappab}
\end{eqnarray}
$\delta$ is the separation of carbon planes in graphite, $f_{\lambda}^{0}$ is the Bose-Einstein distribution for the $\lambda$-branch and $v^{\alpha}_{\lambda}$ the velocity in the $\alpha$-direction.
According to Mathiessen's rule the total phonon lifetime is $1/\tau^{tot}_{\lambda}= 1/\tau^{N}_{\lambda} + 1/\tau^{U}_{\lambda} + 1/\tau^{dis}_{\lambda}$. We remind that the disorder contribution is $\tau^{dis}_{\lambda}=\frac{\hbar}{\Gamma_{\lambda}}$. The full expression of $\Delta\kappa^{\alpha}$ is given in the Appendix.
For the thermal conductivity calculations, we use the $\textit{ab initio}$ results for $\tau^{N}_{\lambda}$ and $\tau^{U}_{\lambda}$ as found in ref.\cite{pauletto}. For $\tau^{dis}_{\lambda}$, we use the analytical forms discussed previously (see details in the Appendix). In Fig.~\ref{fig5}, the average thermal conductivity $\kappa=(\kappa^{x}+\kappa^{y})/2$ (see eq.\ref{kappa}) is plotted as a function of temperature for various concentrations of vacancies. In the clean limit ($x=0$), we find very high values of the thermal conductivity and because of the presence of the U-processes, we observe a 1/T suppression of $\kappa$. As we introduce a very small amount of vacancies of the order of $x=10^{-5}$ only, the effects are dramatic below 200 K. At room temperature, $\kappa$ is already reduced by 40 $\%$ and the suppression is even stronger for $x=10^{-4}$. Indeed, for this concentration, $\kappa=1000$ W.m.$^{-1}$K$^{-1}$ at room temperature, five times smaller than that of pristine graphene. As we increase the vacancy concentration to 0.1$\%$, the thermal conductivity falls to $\kappa=200$ W.m.$^{-1}$K$^{-1}$, 25 times smaller than that of pristine. The variation of $\kappa$ at 300 K is plotted in the inset as a function of $x$ and reveals a crossover around $x=10^{-5}$.
Below this concentration, $\kappa$ is weakly sensitive to the defects concentration, and above, $\kappa$ decreases rapidly with a power law decay $\kappa \propto 1/x^{0.623}$. There is no simple way to anticipate such an exponent.
Let us now compare our results to experimental measurements.\cite{balandin1,nika,chen} At room temperature, in the pristine limit, we obtain $\kappa = 4800$ W.m.$^{-1}$K$^{-1}$ which is in very good agreement with the highest experimental values of the thermal conductivity 
ever reported.\cite{balandin1} The agreement with the other data sets beyond room temperature is also relatively good. The measured values are consistent with an extremely low concentration of defects of the order of $x=10^{-6}$ to 10$^{-5}$ suggesting that the samples should be of good quality. Notice that, the agreement between theory and experiments, in the pristine case, has already been achieved in the full BTE calculations of Refs.\cite{lindsay2,fugallo}. Finally, it is interesting to comment on the validity of the RTA approximation.
The study of $\frac{\Delta\kappa}{\kappa_{RTA}}$ as a function of T (see figure in the Appendix)
has revealed that the correction to RTA is extremely large and even dominant for the lowest concentrations of vacancies, typically when $x\leq$ 10$^{-4}$. However, when the vacancy concentration is large enough beyond 0.1$\%$ of the C atoms, the correction becomes very small.

\section{Conclusion}

Combining state of the art \textit{ab initio} approaches with a full and exact treatment of the disorder, we have addressed the impact of vacancies on phonon thermal transport in graphene. It has been found that the vacancy induced multiple scattering contribution to phonon lifetimes are much larger than predicted by second order perturbation theory. Furthermore, vacancies have drastic effects on both dispersion and lifetime of ZA modes. The ZA dispersion becomes linear in the vicinity of the Brillouin zone center and the vacancy induced scattering rate is linear in energy instead of the quadratic behaviour often assumed. We have also shown that the vacancies have dramatic effects on the thermal conductivity calculated beyond the relaxation time approximation. A mere 0.1$\%$ of vacancies leads to a spectacular 95$\%$ suppression of the thermal conductivity at room temperature. These findings contribute to a better understanding of thermal transport in 2D materials and could be promising for high-efficiency thermoelectric power generation as the high lattice conductivity in natural graphene is the major obstacle. It is also worth noticing that our methodology is very general, it could easily incorporate other features such as extended defects, porosity and nanostructuring.

\section{Appendix}

\subsection{ {\it{ab initio}} calculations of the Interatomic Force Constants} 

The {\it{ab initio}} calculations are performed with the DFT package
SIESTA.~\cite{siesta} The exchange-correlation functional used here
corresponds to the generalized gradient approximation as proposed by
Perdew, Burke \& Ernzerhof.~\cite{perdew} However, the local
density approximation leads to similar results. Troullier-Martin
norm-conserving pseudopotentials~\cite{troullier} are used. The
basis corresponds to a double-$\zeta$-polarized basis optimized with
the simplex tool of the SIESTA package. All the atomic structures were
optimized up to forces less than $10^{-4}$~eV/$\AA$ and to an
hydrostatic pressure smaller than $10$~bar. Finally, a Monkhorst-pack of
$10\times 10\times 1$ k-points is used for the calculations along with
a mesh cutoff of $600$~Ry.

The IFC, $\phi_{\textbf{ij}}^{\alpha\beta}$, were calculated by the finite displacement method. The displacement amplitude was 0.04 Bohr. Note that, we use the local constraint $\phi_{\textbf{ii}}^{\alpha\beta} = -\sum_{\textbf{j}\ne \textbf{i}} \phi_{\textbf{ij}}^{\alpha\beta}$ that ensures that no force results from a global translation of the whole system.

In the presence of a C vacancy, two different supercells have been used to compute the inter atomic force constants.
First, we have considered a large rectangular supercell with dimensions $21.3$~\AA~$\times 24.6$~$\AA$ 
(199 C atoms) and second a smaller diamond shaped supercell that contains 97 atoms.
In both cases, the nearest-neighbours IFC ($\phi_{n-n}^{zz}$) exhibit the same axial symmetry (y-axis) as that of the chosen supercell.
The average value of the $\phi_{n-n}^{zz}$ of the 2-coordinated atoms is -3.03 eV/${\AA}^{2}$ for the largest supercell and 
-2.82 eV/${\AA}^{2}$ for the smallest. The fluctuations around the average, were about 0.8 eV/${\AA}^{2}$ and 0.4 eV/${\AA}^{2}$ respectively for the rectangular and diamond shaped supercells. These fluctuations originates from the non-symmetric free relaxation of the C atoms in the supercells. The angles involving C$_2$ and C$_3$ of Fig. 1 in the manuscript are close to 121$^{\circ}$, in contrast, that of C$_1$ is around 126$^{\circ}$ for the largest supercell, and the distance were slightly shorter for the bonds involving C$_1$.

\subsection{The Chebyshev Polynomials Green's Function} 

In the presence of the C vacancies, the matrix elements of the 3N $\times$ 3N (N total number of C atoms) disordered dynamical matrix $\hat{\bf D}$ read,
\begin{eqnarray}
D^{\alpha\beta}_{\textbf{i}\textbf{j}}=\bra{\textbf{i},\alpha} \hat{\bf D} \ket{\textbf{j},\beta} = \frac{\phi^{\alpha\beta}_{\textbf{i}\textbf{j}}}{\sqrt{m_{\textbf{i}}m_{\textbf{j}}}}x_{\textbf{i}}x_{\textbf{j}},
\end{eqnarray}
where $\alpha=$ x, y and z, $\textbf{i}$ and $\textbf{j}$ are the site positions. The random variable $x_{i}$ is zero if the site is occupied by a vacancy, otherwise 1. In addition for 2-coordinated C atoms and if $\textbf{i}$ and $\textbf{j}$ are nearest-neighbours $\phi^{\alpha\beta}_{\textbf{i}\textbf{j}}$ is half of its value in the pristine compound. We define the plane wave state,
\begin{eqnarray}
\ket{\textbf{q},\alpha}=\frac{1}{\sqrt{N}} \sum_{\textbf{i}} e^{i \textbf{q}.\textbf{r}_{i}} \ket{\textbf{i},\alpha},
\end{eqnarray}
where the sum runs over the sites occupied by C atoms. 
For in plane modes, for a chosen polarisation vector $\textbf{e}_{s}=\cos(\theta_{s}).\textbf{e}_x + \sin(\theta_{s}).\textbf{e}_y$, we define $\ket{\textbf{q},s}=cos(\theta_{s})\ket{\textbf{q},x} + \sin(\theta_{s})\ket{\textbf{q},y}$. Hence, $\textbf{e}_{s}=$ $(\frac{q_{x}}{\vert\textbf{q} \vert},\frac{q_{y}}{\vert \textbf{q} \vert})$, $(\frac{-q_{y}}{\vert\textbf{q} \vert},\frac{q_{x}}{\vert \textbf{q}\vert})$ and $(\frac{1}{\sqrt{2}}, \frac{1}{\sqrt{2}})$ for respectively longitudinal, transverse and (1,1)-axis polarisation.
For the out of plane modes (ZO, ZA) $\textbf{e}_{s}=\textbf{e}_z$. 

To calculate the dynamical spectral function, we fix the polarization and choose a vector $\textbf{q}$ in the discretized BZ. The procedure described in what follows is repeated for a large number of $\textbf{q}$. The phonon dynamical spectral function is given by, 
\begin{eqnarray}
A_{s}(\textbf{q},\omega)=-\frac{2}{\pi} \omega.\Im (G_{s}(\textbf{q},\omega)), 
\label{Aqw2} 
\end{eqnarray}
where the Green's function reads, 
\begin{eqnarray}
G_{s}(\textbf{q},\omega)= \lim_{\eta \to 0}\bra{\textbf{q},s}\Big[z\hat{\bf 1} -\hat{\bf D}\Big]^{-1}\ket{\textbf{q},s},
\label{GFb}
\end{eqnarray}
where $z=\omega^{2}+i\eta$, $G_{s}(\textbf{q},\omega)$ is calculated using the Chebyshev Polynomials Green's Function method (CPGF) as described in details in Refs \cite{mucciolo,weisse}.
The starting point is a rescaling of the dynamical matrix such that the spectrum is contained in the interval [-1, 1], in fact [0, 1] here.
We define the rescaled matrix $\textbf{D}_{r}=\hat{\textbf{D}}/{a}$, $\omega_{r}^{2}=\omega^{2}/a$ and $\eta_{r}=\eta/a$.
The crucial step is the fact that the rescaled Green's function can be expanded exactly in terms of the Chebyshev polynomials 
T$_{n}$, \cite{mucciolo}
\begin{eqnarray}
\Big[z_{r}\hat{\bf 1} -\hat{\bf D_{r}}\Big]^{-1} = \sum^{\infty}_{n=0} g_{n}(z_{r}) T_{n}(\hat{\bf D_{r}}),
\end{eqnarray}
where $z_{r}= \omega_{r}^{2} + i\eta_{r}$,
\begin{eqnarray}
g_{n}(z)=\frac{2i^{-1}}{1+\delta_{n,0}} \frac{(z-i\sqrt{1-z^{2}})^{n}}{\sqrt{1-z^{2}}}.
\end{eqnarray}
$T_{n}(\hat{\bf D}_{r})$ obeys the recursion relation,
\begin{eqnarray} 
T_{n+1}(\hat{\bf D}_{r})= 2\, \hat{\bf D}_{r}.T_{n}(\hat{\bf D}_{r})-T_{n-1}(\hat{\bf D}_{r}),
\end{eqnarray}
with $T_{0}(\hat{\bf D}_{r})=\hat{\bf 1}$ and $T_{1}(\hat{\bf D}_{r})=\hat{\bf D}_{r}$. This implies that eq.(\ref{GFb}) becomes,
\begin{eqnarray}
G_{s}(\textbf{q},\omega)= \lim_{\eta \to 0} \frac{1}{a} \sum^{\infty}_{n=0} \mu_{n}(\textbf{q},s) g_{n}(z_{r}),
\label{GFbn}
\end{eqnarray}
where, 
\begin{eqnarray}
\mu_{n}(\textbf{q},s)= \bra{\textbf{q},s}  T_{n}(\hat{\bf D}_{r}) \ket{\textbf{q},s},
\end{eqnarray}
are called the moments.

In practice, the series of eq.(\ref{GFbn}) is truncated when the convergence for a chosen value of $\eta$ is reached within a certain accuracy.
The efficiency and power of the CPGF method is that the moments $\mu_{n}(\textbf{q},s)$ are obtained iteratively using the recursion relation of the Chebyshev polynomials without any need to store large matrices. Note that, to extract reliably the C vacancy contribution to the phonon lifetime, $\sqrt{\eta}$ has to be much smaller than the full width at half maximum of the peaks in the spectral function. As a consequence, a large number of moments is required as we approach the $\Gamma$ point in the BZ.

\subsection{The Callaway theory} 

The full and detailed derivation of the Callaway method can be found in Refs.\cite{callaway,allen,ma}. Here, we just summarize the main results. The total conductivity in the $\alpha$-direction can be written,
\begin{eqnarray}
\kappa^{\alpha}(T)=\kappa^{\alpha}_{RTA}(T) + \Delta\kappa^{\alpha}(T),
\label{kappac}
\end{eqnarray}
where, the RTA contribution is,
\begin{eqnarray}
\kappa_{RTA}^{\alpha}=\frac{1}{k_{B}T^{2}}\frac{1}{N\Omega\delta} \sum_{\textbf{q},\lambda} (\hbar \omega_{\lambda})^{2}(v^{\alpha}_{\lambda})^{2}\times \nonumber \\
\tau^{tot}_{\lambda}f_{\lambda}^{0}(f_{\lambda}^{0}+1),
\label{kappa-rta}
\end{eqnarray}
N is the total number of unit cells, $\lambda$ the mode index (LA, TA, ZA), $\Omega$ the primitive cell area, $\delta$ the distance between graphene sheets in graphite, $f_{\lambda}^{0}$ is the Bose distribution, $\hbar \omega_{\lambda}$ the mode energy, $v^{\alpha}_{\lambda}$ its velocity in the $\alpha$-direction and $\tau^{tot}_{\lambda}$ the inverse of the total scattering rate for the branch $\lambda$. According to Mathiessen's rule the total phonon lifetime is, 
\begin{eqnarray}
1/\tau^{tot}_{\lambda}= 1/\tau^{N}_{\lambda} + 1/\tau^{U}_{\lambda} + 1/\tau^{dis}_{\lambda}.
\end{eqnarray}
The first step consists in replacing in the standard Boltzmann equation the collision rate by,
\begin{eqnarray}
\left(  \frac{\partial f_{\lambda}}{\partial t}  \right)_{c}=\frac{f^{d}_{\lambda}-f_{\lambda}}{\tau_{N}} + \frac{f^{0}_{\lambda}-f_{\lambda}}{\tau_{R}},
\end{eqnarray}
$f^{d}_{\lambda}$ is the drifted distribution function, and we have defined the scattering rate for resistive processes,
$1/\tau^{R}_{\lambda}= 1/\tau^{U}_{\lambda} + 1/\tau^{dis}_{\lambda}$. Only the resistive processes tend to bring 
$f_{\lambda}$ back to its equilibrium value. The drifted distribution $f^{d}_{\lambda}$ is defined by
\begin{eqnarray}
f^{d}_{\lambda}=\frac{1}{e^{\beta(\omega_{\lambda}(\textbf{q})-\textbf{v}_{d}.\textbf{q})}-1},
\end{eqnarray}
where we have introduced as a Lagrange multiplier the drift velocity $\textbf{v}_{d}$. This quantity is determined by the condition that the normal processes conserve the momentum. By following step by step Callaway's derivation we find,
\begin{eqnarray}
\Delta\kappa^{\alpha}(T)= \frac{(A^{\alpha}(T))^{2}}{B^{\alpha}(T)},
\label{deltakappa}
\end{eqnarray}
where the numerator $A^{\alpha}(T)=\sum_{\lambda} A^{\alpha}_{\lambda}(T)$ and,
\begin{eqnarray}
A^{\alpha}_{\lambda}(T)= \frac{1}{k_{B}T^{2}}\frac{1}{N\Omega\delta} \sum_{\textbf{q}} \frac{\tau^{tot}_{\lambda}}{\tau^{N}_{\lambda}}  \hbar \omega_{\lambda} v^{\alpha}_{\lambda} q^{\alpha}_{\lambda} \times \nonumber \\ 
\;\;\;\;f_{\lambda}^{0}(f_{\lambda}^{0}+1),
\end{eqnarray}

and the denominator $B^{\alpha}(T)=\sum_{\lambda} B^{\alpha}_{\lambda}(T)$ and,
\begin{eqnarray}
B^{\alpha}_{\lambda}(T)=\frac{1}{k_{B}T^{2}}\frac{1}{N\Omega\delta} \sum_{\textbf{q}}  \frac{\tau^{tot}_{\lambda}}{\tau^{N}_{\lambda} \tau^{R}_{\lambda}} \times \nonumber \\
(q^{\alpha}_{\lambda})^{2} f_{\lambda}^{0}(f_{\lambda}^{0}+1).
\end{eqnarray}

\begin{figure}[t]\centerline
{\includegraphics[width=1.\linewidth,angle=0]{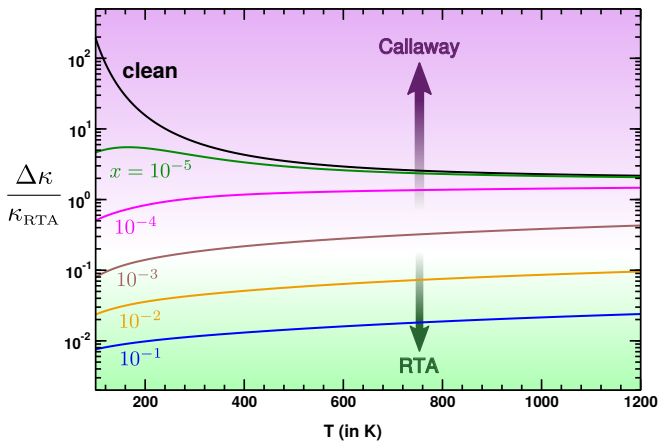}}
\caption{Callaway's correction to the total thermal conductivity over RTA contribution as a function of temperature, for various concentrations of C vacancies. The green area corresponds to negligible Callaway correction ("RTA regime") and the purple one to dominant Callaway correction.
}
\label{fig6}
\end{figure} 

To facilitate the calculations we replace the discrete sum over $\textbf{q}$ by an integral over energy and thus we introduce Debye frequencies for LA, TA and ZA modes. This leads to,
\begin{eqnarray}
A^{\alpha}_{\lambda}(T)= c_{\lambda}\frac{1}{\Omega\delta} \int^{\omega_{D\lambda}}_{0} \frac{\tau^{tot}_{\lambda}}{\tau^{N}_{\lambda}} \omega  \rho_{\lambda}(\omega) \frac{\partial f^{0}}{\partial T} d\omega,
\label{aa}
\end{eqnarray}
where the coefficient $c_{\lambda}$=1/2 for LA and TA branches and 1 for ZA.
Similarly $B^{\alpha}_{\lambda}(T)$ can be rewritten,
\begin{eqnarray}
B^{\alpha}_{\lambda}(T)= \frac{1}{2\hbar\Omega\delta} \int^{\omega_{D\lambda}}_{0} \frac{\tau^{tot}_{\lambda}}{\tau^{N}_{\lambda}\tau^{R}_{\lambda}} g_{\lambda}(\omega) \times \nonumber \\ 
\rho_{\lambda}(\omega)\frac{\partial f^{0}}{\partial T} d\omega.
\label{bb}
\end{eqnarray}
The density of states is $\rho_{\lambda}(\omega) = \frac{\Omega}{2\pi v^{2}_{\lambda}} \omega$ for both LA and TA, and $\frac{\Omega}{4\pi D}$ for ZA. $v_{\lambda}$ is the velocity at the $\Gamma$ point for LA and TA, and D is the stiffness of the ZA branch.
$g_{\lambda}(\omega)= \frac{\omega}{v^{2}_{\lambda}}$ for LA and TA, and 1/D for ZA. Note that, in eq.(\ref{aa}) and (\ref{bb}) we have used the linear dispersion for LA and TA and the quadratic one for ZA.

To compute numerically the total conductivity as a function of temperature, we need the temperature dependence of $\tau^{N}_{\lambda}$ and $\tau^{U}_{\lambda}$. We use the $\textit{ab initio}$ calculated scattering rate at room temperature (T$_{0}$) for both U and N processes ($\tau^{N}_{0,\lambda}$, $\tau^{U}_{0,\lambda}$) extracted from Ref.\cite{pauletto}.
We make the following ansatz to have the full T variation of the scattering rates,
\begin{eqnarray}
1/\tau^{N}_{\lambda} (T)=(1/\tau^{N}_{0,\lambda}) \frac{T}{T_{0}}
\end{eqnarray}
and the Umklapp scattering rate is,
\begin{eqnarray}
1/\tau^{U}_{\lambda} (T)=(1/\tau^{U}_{0,\lambda}) \frac{T}{T_{0}} e^{-\frac{\omega_{D\lambda}}{3} (\frac{1}{T} - \frac{1}{T_{0}})}. 
\end{eqnarray}
This ansatz is motivated by the fact that the form usually assumed for the N and U process are,
$1/\tau^{N}_{\lambda}= \omega^{a} T^{b}$ and $1/\tau^{U}_{\lambda} \propto (1/\tau^{N}_{\lambda}) e^{-\frac{\omega_{D\lambda}}{cT}}$ 
where a = 1 or 2, b = 1, 2 or 3 and c is often set to 3. \cite{holland,morelli,srivastava,callaway} 
For the vacancy concentration dependent scattering $1/\tau^{dis}_{\lambda}$ we use the results found in the present study.

We now discuss the importance of Callaway's correction  to the thermal conductivity in graphene, in the presence of C vacancies. The results are  depicted in Fig. \ref{fig6}. Let us first consider low vacancy concentrations ($x\leq 10^{-4}$). For this range of concentration, the correction $\Delta\kappa^{\alpha}(T)$ is important and even dominates. For instance the ratio $\Delta\kappa/\kappa_{RTA}$ is 1, 4.5 and 7.5 for $x=10^{-4}$, 
10$^{-5}$ and for the pristine case at room temperature. On the other hand, when the vacancy concentration is large enough beyond 0.1$\%$  of the C atoms, the correction becomes very small, it is less than 15$\%$ of the RTA value. For $x=$ 1$\%$ the correction to RTA represents only 5$\%$ of the total conductivity.

These results show that in the presence of a sufficient amount of C vacancies, beyond 0.1$\%$, the RTA approach becomes appropriate for the evaluation of the thermal conductivity, provided that the scattering rates are calculated accurately.

\end{document}